\documentclass{emulateapj}

\usepackage[latin1]{inputenc}
\usepackage[english]{babel}
\usepackage{amsmath}
\usepackage{amssymb}		
\usepackage{latexsym} 
\usepackage{epsfig}
\usepackage{subfigure}
\usepackage{natbib}
\usepackage{setspace}
\usepackage{graphicx}
\usepackage{longtable}
\newcommand{\msun}{{M_{\odot}}}
\newcommand{\mstar}{{M_{\ast}}}
\newcommand{\ser}{S\'ersic }

\shorttitle{Evolution of the mass-size relation to $z = 3.5$ in the GOODS North field}
\shortauthors{Mosleh et al.}

\begin{document}

\title{The Evolution of the Mass-size relation to z=3.5 for UV-bright galaxies and sub-mm galaxies in the GOODS-NORTH field}

\author{Moein Mosleh\altaffilmark{1}, Rik J. Williams \altaffilmark{2}, Marijn Franx \altaffilmark{1}, Mariska Kriek\altaffilmark{3}} \altaffiltext{1}{Leiden Observatory, Universiteit Leiden, 2300 RA Leiden, The Netherlands}\email{mosleh@strw.leidenuniv.nl}
\altaffiltext{2}{Carnegie Observatory, Pasadena, CA 91101, USA}
\altaffiltext{3}{Harvard-Smithsonian Center for Astrophysics, 60 Garden Street, Cambridge, MA 02138, USA}

\begin{abstract}

We study the evolution of the size - stellar mass relation for a large spectroscopic sample of galaxies in the GOODs North field up to $z \sim 3.5$. The sizes of the galaxies are measured from $\textit{K}_{s}$-band images (corresponding to rest-frame optical/NIR) from the Subaru 8m telescope. We reproduce earlier results based on photometric redshifts that the sizes of galaxies at a given mass evolve with redshift. Specifically, we compare sizes of UV-bright galaxies at a range of redshifts: Lyman break galaxies (LBGs) selected through the U-drop technique ($z \sim 2.5-3.5$), BM/BX galaxies at $z \sim 1.5-2.5$, and GALEX LBGs at low redshift ($z \sim 0.6-1.5$). The median sizes of these UV-bright galaxies evolve as $(1+z)^{-1.11\pm0.13}$ between $z \sim 0.5-3.5$.  The UV-bright galaxies are significantly larger than quiescent galaxies at the same mass and redshift by $0.45\pm0.09$ dex. We also verify the correlation between color and stellar mass density of galaxies to high redshifts. The sizes of sub-mm galaxies in the same field are measured and compared with BM/BX galaxies. We find that median half-light radii of SMGs is $2.90 \pm 0.45$ kpc and there is little difference in their size distribution to the UV-bright star forming galaxies.

\end{abstract}


\section{Introduction}

Recent studies provide evidence that sizes of galaxies at high redshifts were smaller in comparison with galaxies of similar mass in the local universe \citep[e.g.][]{daddi2005,trujillo2006b,trujillo2006a,zirm2007,toft2007,dokkum2008, franx2008, williams2009, toft2009, Damjanov2009, Cimatti2008, longhetti2007}. These studies pose several questions about the evolution of properties of galaxies and the build up of galaxy stellar mass with time. Several physical processes are proposed to explain the growth of galaxies with time, such as galaxy (major or minor) mergers \citep[e.g.][]{khochfar2006,khochfar2009, bell2006, naab2009}, gas accretion in outer regions and star formation \citep[e.g.][]{dekel2009,Elmegreen2008}. However, each of these mechanisms will affect differently the growth of mass and size of galaxies. For example, major dry mergers could produce descendants with larger than observed stellar masses for early-type galaxies. Therefore, some authors \citep[e.g.][]{dokkum2010, bezanson2009, naab2009, hopkins2009} suggest minor mergers and envelope growth by satellite accretion to be a more likely mechanism for building up quiescent galaxies. In general, plausible models have to reproduce galaxy sizes and stellar mass densities seen today and bring the high redshift galaxies to the local stellar mass-size relation. Therefore, exploring the observed evolution of galaxy sizes is essential to constrain these galaxy formation and evolution models. 

However, most of the size studies at high redshift were based on photometric redshifts. Although techniques to construct galaxy spectral energy distributions (SEDs) from multiwavelength observations and photometric redshift measurements are improved over the past few years, photometric redshifts still have worse uncertainties for star forming galaxies due to the lack of a strong 4000\AA \ break. The large uncertainties and possible systematic biases, due to the reliance on photometric redshifts can seriously affect the masses and star-formation rate measurements and thus the final results. Spectroscopic observations of high-redshift galaxies and making complete magnitude limited samples require large amounts of observing time. Nevertheless, the advent of 8-10m class telescopes with multi object spectroscopy has brought about much larger high-$z$ spectroscopic samples than were previously  available. 

Spectroscopic redshifts are relatively easier to obtain for star forming galaxies (compare to quiescent galaxies) due to emission lines. Previous results on the size evolution of this population (with photometric redshifts) are interesting since some of the usual mechanisms assumed for size growth, e.g., gas-poor (``dry'') mergers do not apply to gas-rich starburst galaxies since substantial amounts of gas is required for these galaxies to undergo significant star formation. Therefore, studies of the structure of star forming galaxies with secure redshifts can provide strong constraints on size evolution models. 

The actively star forming galaxies at high redshift consists of two main populations: relatively unobscured UV-bright galaxies and dusty red starbursts, the most extreme of which are detected as sub-millimetre galaxies (SMGs). The UV-bright galaxies are selected to have strong rest-frame UV. In the past few years, many simple photometric techniques (e.g., ``\textit{U}-dropout'') were designed to select these galaxies at different redshift ranges \citep[e.g.][]{steidel2003, adelberger2004}. 

The SMGs are among the most massive, luminous, and vigorously star forming galaxies at high redshift that are heavily obscured by dust \citep[e.g.][]{hughes1998,smail2002,chapmannatur, chapman2005}. The physical process driving these highly luminous galaxies (star-formation, AGN or combination of these two) is still uncertain. The evolution of SMGs and their relation to the local galaxies are also not yet understood: for instance, whether or not SMGs are the progenitors of local elliptical galaxies \citep[e.g.][]{blain2004, swinbank2006, tacconi2008}. Their position on the size-mass plane may therefore provide clues about their relation to other galaxy populations.

To verify previous results based on photometric redshifts, here we study the mass and size evolution of a large sample of UV-bright
and submillimetre galaxies in GOODS-North with secure spectroscopic redshifts. The structure of this paper is as follows. In section 2, we review the data. In section 3 \& 4, we describe size and mass determinations of galaxies. Our sub-samples selection are described in section 5. Finally, in section 6 we present our results and investigate the size evolution and stellar mass-size relation for our samples. We summarize and discuss our results in section 7.  The cosmological parameters adopted throughout this paper are $\Omega_{m}$ = 0.3, $\Omega_{\Lambda}$ = 0.7 and $H_{0} = 70$ $km$ $s^{-1}$ $Mpc^{-1}$.\\

\section{Description of Data}

The sample of galaxies used here is based on the most complete spectroscopic catalog of galaxies in the GOODS (Great Observatories Origin Deep Survey \citep{giavalisco2004}) North field by \cite{barger2008}. This catalog gives a compilation of all spectroscopic observations carried out in this field \citep[e.g.][]{cowie2004, reddy2006, wirth2004, cohen2001, cohen2000}, where each galaxy sample was selected in a different way. In addition, \cite{barger2008} performed spectroscopic observations for certain subsamples. The catalog includes 2907 sources including stars and is restricted to sources with $K_{s, AB} < 24.5 $ or $F850LP_{AB} < 26$. There are 2362 sources with $ z < 1.6 $ and 327 sources between $ z =$ 1.6 and 3.5. Most redshifts for galaxies with $z > 1.6$ come from \cite{reddy2006}. 

This catalog \citep{barger2008} comprises optical photometric data in the F435W, F606W, F775W and F850LP passbands taken from HST Advanced Camera for Surveys (ACS) \citep{giavalisco2004} and the U-band magnitude is taken from \cite{capak2004}. The near-infrared $K_{s}$-band magnitude measured from WIRCam images from CFHT. Details of near-IR observations, data reduction and generating catalog are described more in \cite{barger2008}. X-ray soft ($0.5-2 KeV$) \& hard ($2-8 KeV$) luminosities are also provided for many sources in the catalog. In \cite{barger2008} sources with X-ray luminosities above $ 10^{42}$ $ergs $ $ s^{-1}$ in either soft or hard band defined as AGNs. For a fraction of sources ($\sim 36\%$) near-ultraviolet (NUV) and far-ultraviolet (FUV) magnitudes from the UV imaging survey performed by GALEX mission were also provided. Note that sources indicated as stars in the catalog are excluded from our analysis.

\section{Sizes}

\begin{figure*}
\includegraphics[width=\textwidth]{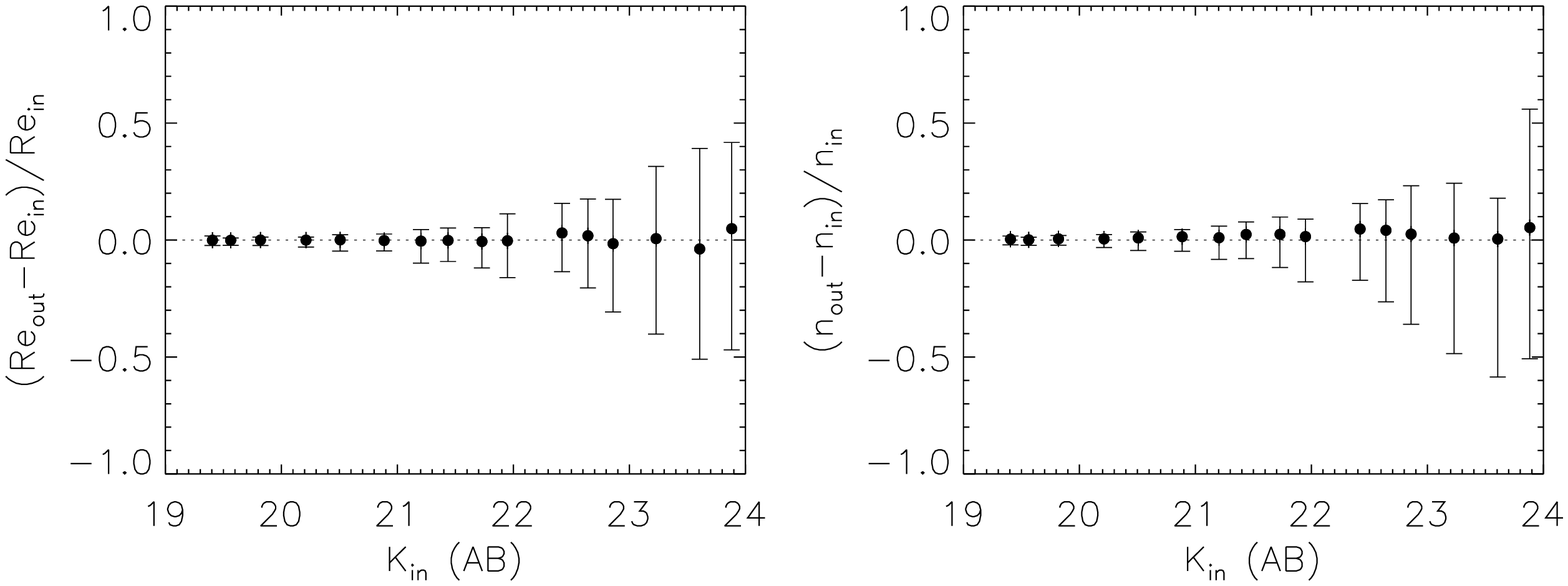}
\caption{\textit{Left panel}: The points are the median of the relative difference between the recovered and input sizes versus magnitude based on our simulations. \textit{Right panel}: The same comparison but for the relative difference between output and input \ser index. The error bars illustrate the $68 \%$ scatter. The random uncertainties of recovered sizes and \ser index increase with magnitude.}
\label{fig1}
\end{figure*} 

\begin{figure}
\includegraphics[width=3.4 in]{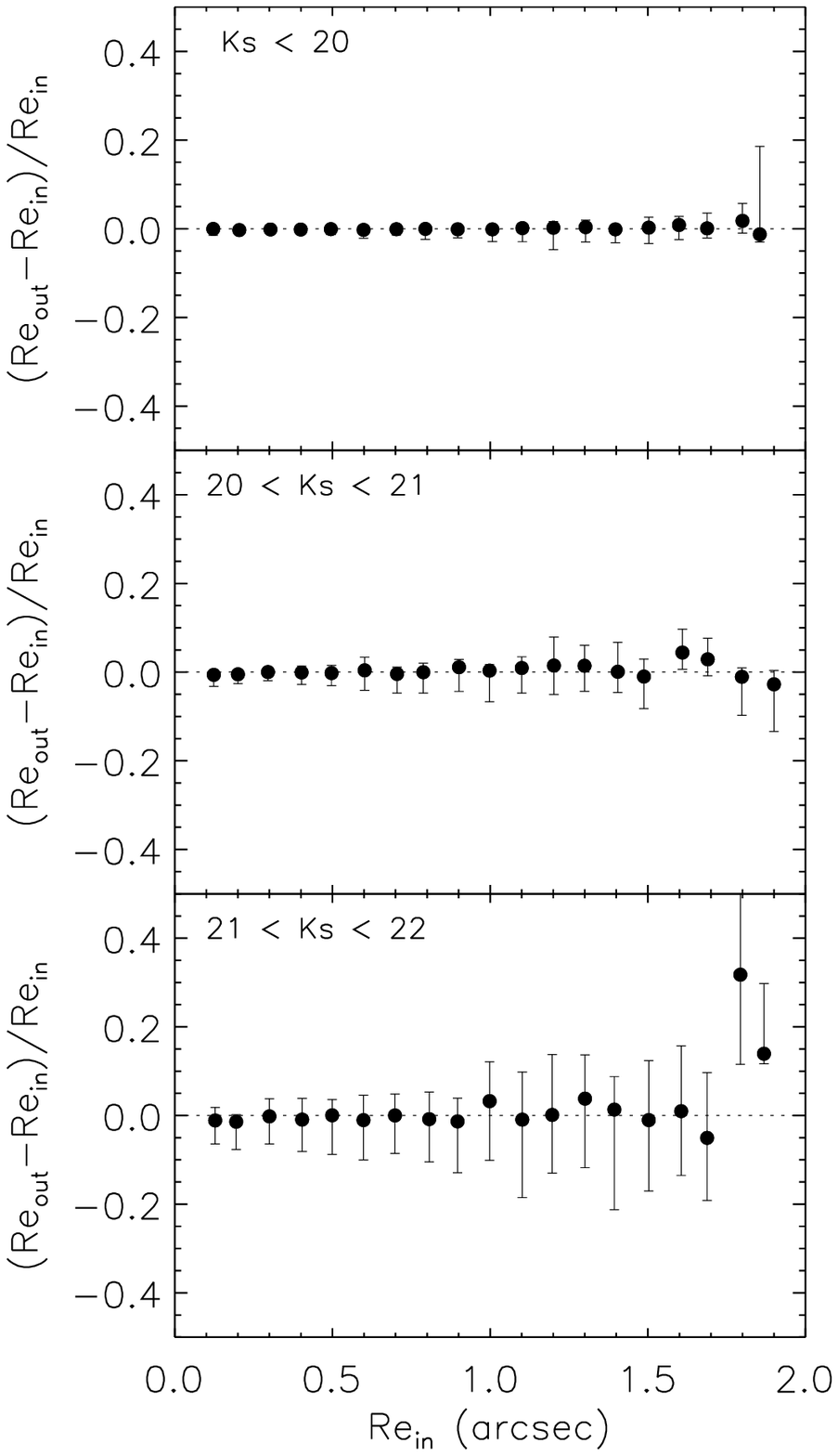}
\caption{Relative difference between the input and measured sizes of galaxies in our simulations as a function of size for mock galaxies for galaxies with $K_{AB} \leq 20$ mag (top panel), $20 < K_{AB} < 21$ (middle panel) and $21 < K_{AB} < 22$ (lower panel). The relative errors between the input and measured sizes of galaxies as a function of sizes depends on the magnitude. }
\label{fig2}
\end{figure} 

\subsection{Size measurements}
Size determination in the observed near-infrared (rest-frame optical at high $z$) is more robust in tracing the distribution of stellar mass than the observed optical (rest-frame UV at high $z$) since the sizes at rest-frame UV can be strongly affected by dust and star-formation. Therefore, we use $\textit{K}_{s}$-band images from Subaru MOIRCS Deep Survey (MODS) in the GOODS North field \citep{kajisawa2006,ouchi2007}. The imaging observations are performed for $J$ and $K_{s}$ bands over $\sim112$ arcmin$^{2}$ area of the GOODS-N field with a pixel scale of 0.12 arcsec. We used the images reduced by \cite{bouwens08}. The resulting FWHM of the $K_{s}$ images is $\sim0.5''$. The deepest data cover $\sim28$ arcmin$^{2}$ reaching a 5 $\sigma$ of 25.4 AB mag in $K_{s}$ band while the other regions are $\sim 1.3$ mag shallower. More details of the images and data reduction can be found in \cite{bouwens08}.

 Sizes of galaxies were estimated by using the GALFIT code \citep{peng2002} in a similar procedure used by \cite{trujillo2007} and \cite{williams2009}. For each galaxy, a square postage stamp of 120 pixels (14.4'') around each galaxy was made and we use a mask to exclude neighbouring galaxies from the fit. A range of \ser (1968) profile models convolved with the point-spread function (PSF) of the image are fitted to the $\textit{K}_{s}$ images of galaxies. The convolved models for each object were compared with the galaxy surface brightness distribution and the best fit model determined using minimized $\chi ^{2}$ of the fit. The PSF used by GALFIT was taken from the median of the unsaturated stars over the entire MOIRCS $\textit{K}_{s}$ images. We perform a test to check the PSF of the image by measuring sizes of a number of stars in the field listed in the \cite{barger2008} catalog. The derived sizes of all the stars are found to be $\ll 0.02$ pixel (effectively zero). Therefore, the PSF is a good approximation of a point source. We note that within our $\textit{K}_{s}$ band images, there is a shallower region with seeing of 0.68''. We followed a similar procedure with an appropriate PSF (which is made from the median of stars in this region) to measure sizes of galaxies in this region separately. We also verified that the derived sizes of the shallower region are reliable by its appropriate PSF.  

We determine the circularized effective radius $r_{e} = a_{e} \sqrt{(1-\epsilon)}$, from the half-light radius along the semimajor axis $a_{e}$ and ellipticity ($\epsilon$) as output by GALFIT. This removes the effects of ellipticity. We allowed the \ser index \textit{n} (which measures the shape of the surface brightness profile of galaxy) to vary between 0.5 and 5 and the effective radius between 0.01 and 60 pixels (0.0012'' and 7.2''). Initial guesses for the effective radius $r_{e}$, ellipticity and position angle were taken from the SExtractor catalog and magnitude was taken from the original \citep{barger2008} catalog, and we set the \ser index to 2 initially. 

Our results show that the median \ser index measured for all galaxies is 2.0 where $50\%$ of measurements lies between 4.0 and 1.0. For galaxies with stellar masses between $10^{10}$-$10^{11} \msun$ the median \ser index is 2.4 where $50\%$ lies between 4.6 and 1.2. 

Due to possibility of color gradients, it is best to use the same rest-frame band for measuring sizes of galaxies at all redshifts. However, only deep $K$ imaging is available, so there is a possibility of systematic effects with redshift. \cite{franx2008} show that such systematic effects are small and will not significantly affect the results. As an approximation, we corrected sizes to the rest-frame $g$ band using star forming galaxies from studies of CDF-South \citep{franx2008}. We derive the best linear fit to the median of the ratio of sizes in $K$ band and rest-frame $g$ band as a function of redshift and apply this fit ($r_{e, k-band}/r_{e, g-band} = 0.15 z + 0.64$) to galaxies with $z \lesssim 2 $ in our sample.

\subsection{Simulations}

To gauge the accuracy of the measured sizes and reliability of our results we perform a realistic simulation. About 8500 \ser profiles were generated with uniformly-distributed random parameters in the ranges of $19 < K_{AB} < 25$, $0.1'' < r_{e} < 2'' $, $0.5 < n < 5$ and $0 < \epsilon < 0.8 $. The mock galaxies were then convolved with the PSF of the image. Finally we added these galaxies to 14.4'' blank-sky postage stamps randomly taken from our $\textit{K}_{s}$ band image. The structural parameters of the model galaxies were then measured in a manner identical to that used for the actual images.

The results of these simulations are shown in Figure 1 and 2. Left panel of Figure \ref{fig1} shows that the scatter in the recovered sizes increase with magnitude. The random uncertainties increase significantly after $K_{AB} = 23 $. However, the systematic errors are very small ($< 5 \%$) even at the faintest magnitudes. Moreover, the simulation shows that the systematic errors on recovery of \ser index are also very small and random uncertainties increase above K =23 (right panel of Figure \ref{fig1}). From this result, we limit the sample studied in this paper to galaxies with $K \leq 23 $ mag. We note that this limit is for the deeper region in our $K$ band images. Running a separate simulation for the shallower region, we find that the derived sizes of galaxies with $K < 22.7 $ mag are reliable. Hence, we restricted the sample in this region to this slightly brighter magnitude limit. 

We have also explored how the recovery of size depends on the size itself. As shown in Figure \ref{fig2}, the uncertainties in retrieving sizes depend on both magnitude and size. Systematic offsets at all magnitudes appear to be negligible except for $r_{e} > 1.8$ arcsec (which corresponds to 14.4 kpc at $z\sim1$) at the faintest magnitudes; however, galaxies this large and faint have not been seen at any redshift. 

The random uncertainties increase with size in all magnitude bins. For objects with $K_{AB} \leq 20$ mag (top panel) the random uncertainties in size recovery are $<5\%$. However, for galaxies with $21 < K_{AB} < 22$ the increasing of random uncertainties with sizes is significant. The increase in random uncertainties at larger sizes could be due to decreasing of surface brightness. However, since we are mainly concerned about the overall sample properties, the lack of systematic errors is more important.

\section{Stellar mass estimates}
The stellar masses of galaxies were measured with the Fitting and Assessment of Synthetic Templates(FAST) code \citep{Kriek2009}. All fluxes from \cite{barger2008} including four optical bands from ACS, the $K_{s}$ infrared and U band were used to find the best-fit galaxy template SED to the broadband photometry using a $\chi ^{2}$ minimization procedure. \cite{BC2003} stellar population evolution models with exponentially declining star formation histories (with $\tau$ ranging from $10^{7-10}$ yr) were used to fit the SEDs. We use the \cite{salpeter1955} initial mass function (IMF) and solar metallicity and the extinction $A_{V}$ was allowed to vary between 0 and 3. Redshift of galaxies were fixed to their spectroscopic redshift provided by \cite{barger2008} catalog. Masses were then shifted by $-0.2$ dex for consistency with the $z \sim 0$ SDSS masses (which were calculated using a \cite{kroupa2001} IMF).\\

\section{Sub-samples}

The ability to identify and study distant galaxies has improved dramatically during the last two decades. Various selection criteria have been designed to select high redshift galaxies through their observed colors. In this paper we use such criteria to select samples of UV-bright galaxies in redshift ranges of $0.6 \leq z \leq 1.4 $, $1.4 \leq z \leq 2.5$ and $ 2.7 \leq z \leq 3.5$.   

Our sample of Lyman break galaxies ($2.7 \leq z \leq 3.5$) is taken from \cite{reddy2006}. Their LBGs candidates were originally preselected by the ``C'', ``D'' and  ``MD'' criteria which use regions of ($U_{n}-G$) versus ($G-R$) color space \citep{steidel2003}. Limiting our sample to $K_{AB} \leq 23 $, we were only able to include 5 LBGs in our analysis. 

At $z \sim 2$, BM/BX galaxies also are taken from \cite{reddy2006}. The BM/BX criteria \citep{adelberger2004, steidel2004} were designed to find actively star forming galaxies at redshifts $1.4 < z < 2.5$  with similar SEDs to the LBGs. The selection criteria are based on the observed $U_{n}GR$ colors of galaxies. \cite{reddy2006} provided spectroscopic observations for candidates brighter than $R_{AB} = 25.5$, photometrically selected from \cite{steidel2003, steidel2004}. For this paper, 41 BM/BX galaxies to $K_{AB} = 23$ are selected from their sample with spectroscopic redshift $z > 1.4$.

Galaxies analogous to the LBGs at $0.6 \leq z \leq 1.4$ (hereafter GALEX/LBGs) are selected with the GALEX/HST $(FUV-NUV)_{AB}$ versus $(NUV-F435W)_{AB}$ color-color diagram. Following the selection criteria used in \cite{barger2008} and requiring  $z > 0.6$,  we include 105 GALEX/LBGs with $K_{AB} \leq 23 $.

\begin{table}\footnotesize
\caption{Sub-samples}
\centering
\label{tb1}
\begin{tabular*}{3.4 in}{c c c c}

  \hline
 \hline
 Sample       & Redshift        & $log(\mstar/\msun)$ & No. of Sources \\  \hline
 GALEX/LBG    & $0.6 < z < 1.4$ &    8.8 - 11.0       & 105 \\
 BM/BX        & $1.4 < z < 2.7$ &    9.8 - 10.8       & 41  \\
 LBG          & $2.7 < z < 3.5$ &    10.2 - 10.8      & 5   \\
 SMG          & $0.5 < z < 3. $ &    10.0 - 11.7      & 14  \\
 \textit{sBzK}& $1.4 < z < 3.2$ &    9.8 - 11.7       & 70  \\
 \hline 
\end{tabular*}
\footnotetext{Samples of star-forming galaxies studied in this paper.}
\end{table}

In addition to the UV-bright galaxies in GOODS-N field, we also study sizes of the Sub-millimetre galaxies (SMGs) as a population of star forming galaxies at high redshifts. The subsample of SMGs for our studies is drawn from \cite{chapman2005} and \cite{pope2006}. The catalog of SMGs provided by \cite{pope2006} contains 35 candidates from GOODS-N field with 21 secure optical counterparts. Of these, only 8 galaxies in our observed images have spectroscopic redshifts. We further include 6 SMGs from the \cite{chapman2005} HDF-North study. Therefore, our final SMG sample studied here contains 14 sources spanning $0.5 \leq z \leq 3$. 

Besides these two populations, star forming galaxies at $z > 1.4$ can also be identified using \textit{BzK} technique \citep{daddi2004}. This technique is designed to cull galaxies from $K$-selected samples. Star forming \textit{BzK} galaxies (\textit{sBzK}) tend to be more massive and have higher reddening than UV-selected galaxies. With this method, 70 \textit{sBzK} galaxies at $z > 1.4$ and $ K_{AB} \leq 23$ are selected using the criteria $(z - K) - (B - z) > -0.2$. We further use this sample to compare properties of star forming galaxy populations at high redshifts. The \textit{sBzK} selection criteria can identifies $\sim90\%$ of our BM/BX galaxies. The large overlap is not surprising since both our UV-selected and \textit{sBzK} samples consist of mostly massive galaxies (due to the $K$ magnitude limit). Table \ref{tb1} lists samples of star forming galaxies used in this paper.

\begin{figure}
\includegraphics[width=3.4 in]{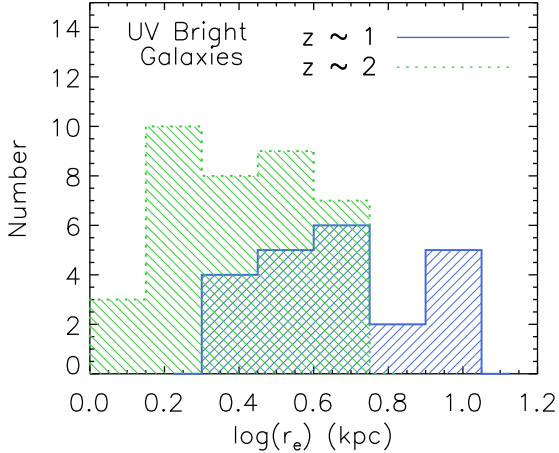}
\caption{Size distribution of UV-bright galaxies with stellar mass $10 < log(\mstar/\msun) < 11$ at $z \sim 1$ (blue histogram) and $z \sim 2$ (green histogram). The half light radii of UV-bright galaxies (GALEX(LBGs)) at $z \sim 1$ are larger than BM/BX galaxies at $z \sim 2$.}
\label{fig3}
\end{figure}

\begin{figure*}
\includegraphics[width=\textwidth]{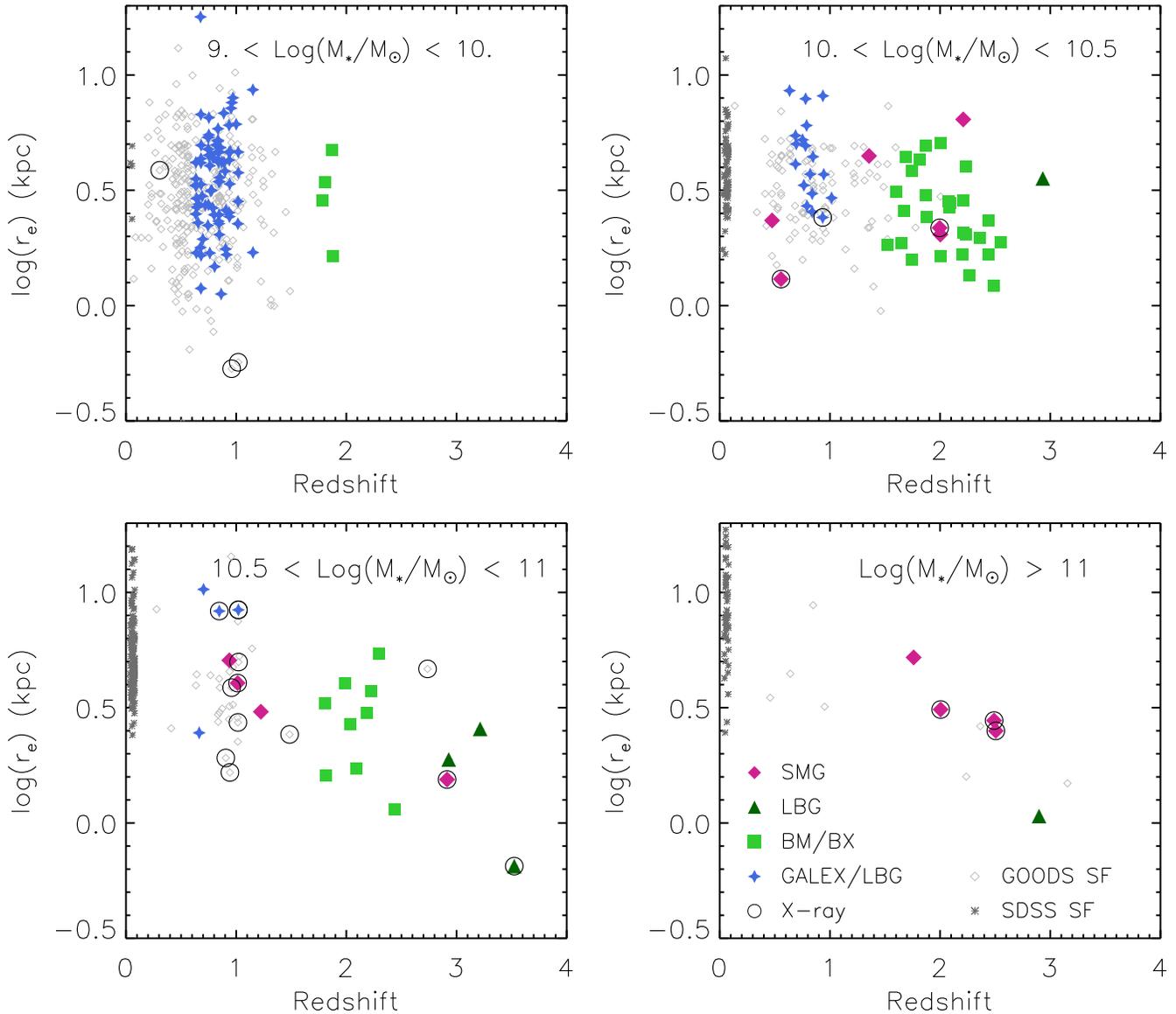}
\caption{Evolution of sizes of galaxies as a function of redshift. Each panel shows the size evolution for a narrow stellar mass bin. The gray symbols are star forming galaxies with spectroscopic redshifts in GOODS-N field. Blue stars are GALEX/LBGs ($ z \sim 1$) while the green squares and green triangles represent BM/BX galaxies ($z \sim 2$) and LBGs ($z \sim 3$)respectively. Submm galaxies are also shown as pink diamonds. Galaxies with X-ray luminosities more than $ 10^{42}$ $ergs $ $ s^{-1}$ are marked by open circles. Black symbols at $z \sim 0$ are SDSS star forming galaxies from \cite{guo2009}.  As this plot illustrates, the half light radii of UV-bright galaxies evolve with redshift and this evolution is faster for high mass galaxies.}
\label{fig4}
\end{figure*}

\section{Results}

\subsection{Size Evolution}

\begin{figure*}
\includegraphics[width=\textwidth]{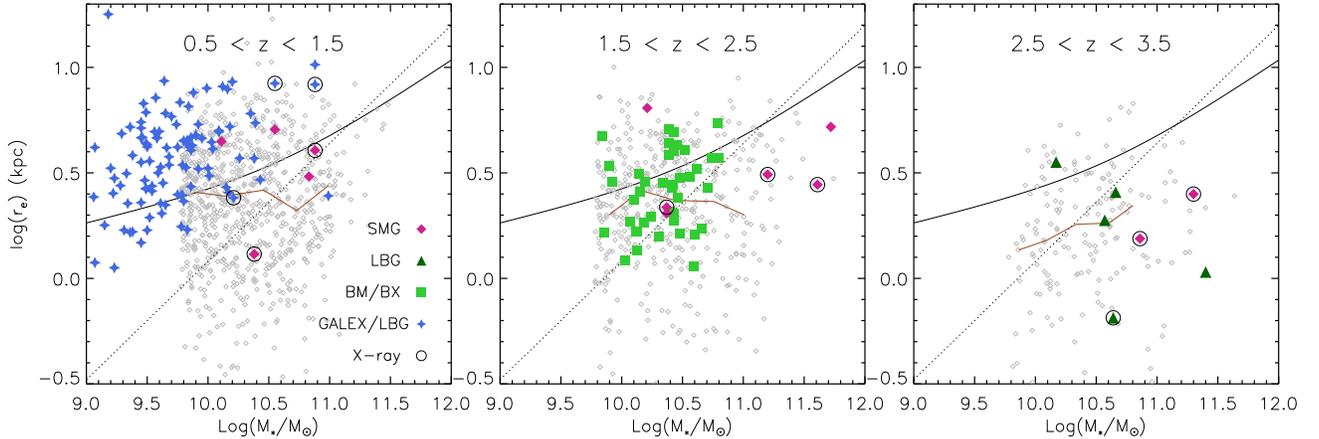}
\caption{Stellar mass size distribution for UV-bright galaxies in different redshift bins compared to the galaxies from CDF-south \citep{franx2008} (gray dots). The color symbols represented here are the same as Fig. \ref{fig4}. The solid and dotted black lines are the size-mass relations for star forming and quiescent galaxies respectively at $z \sim 0$ from \cite{shen2003}. The solid brown lines show the median sizes in narrow mass bins for all galaxies from CDF-S. The UV bright galaxies are in general larger than normal field galaxies at the same mass.}
\label{fig5}
\end{figure*}

\begin{figure*}
\includegraphics[width=\textwidth]{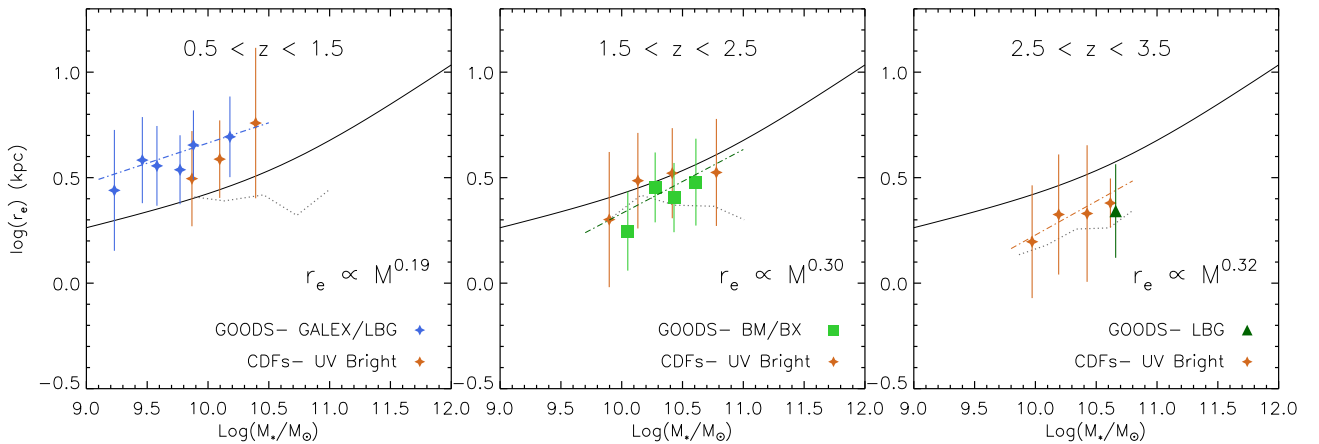}
\caption{Comparison of the stellar mass-size relation for the UV-bright galaxies from GOODS-N field and CDF-South in three redshift bins. The black solid lines are the size-mass relation for star forming galaxies at $z \sim 0$ from \cite{shen2003}. The color symbols are median of UV-bright galaxies in narrow mass bins and the error bars show one $\sigma$ (68\%) dispersion. The dashed-dotted lines are the best power law fits using $r_{e} \propto M^{\alpha}$ to the individual UV-bright galaxies. The dotted lines are the median sizes in narrow mass bins for all galaxies from CDF-S (Same as Fig \ref{fig5}). This plot shows that the derived half-light radii of UV-bright galaxies from both fields are in place and agreement with each other up to $z \sim 3$. }
\label{fig6}
\end{figure*}

\begin{figure}
\includegraphics[width=3.4 in]{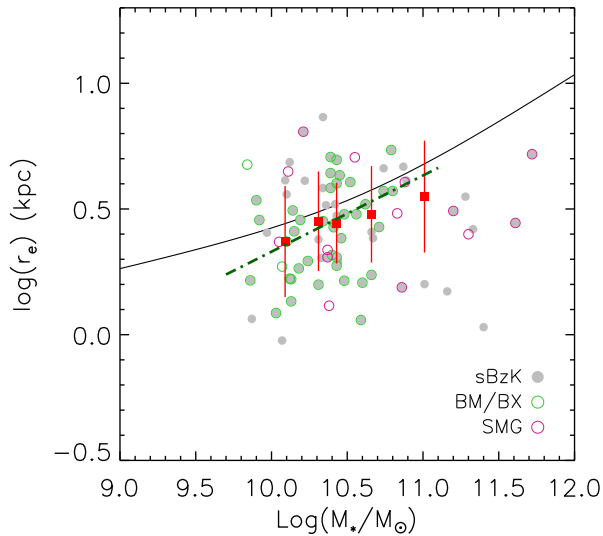}
\caption{Stellar mass size relation for star forming \textit{sBzK} galaxies (gray circles). The BM/BX galaxies and SMGs are marked with green and pink circles respectively. The red squares are the median for the mass-size relation of \textit{sBzK} sample with one $\sigma$ dispersion. The green dashed-dotted line is the best fit to the BM/BX galaxies. The \textit{sBzK} galaxies have a similar size-mass relation to the BM/BX galaxies. The solid line shows the mass-size relation for star-forming galaxies at $z \sim 0$ from \cite{shen2003}.}
\label{fig7}
\end{figure}

\begin{figure*}
\includegraphics[width=\textwidth]{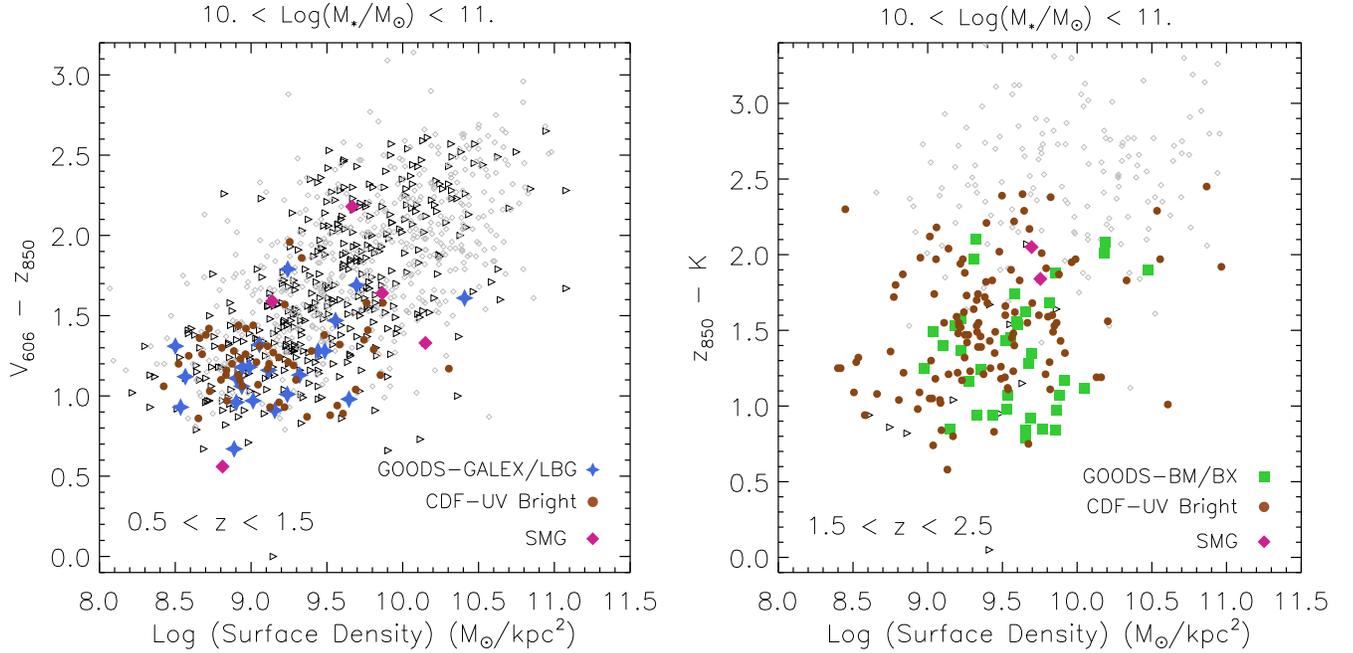}
\caption{\textit{Left panel}: The correlation between color ($V_{606} - z_{850}$) and surface density for galaxies in the redshift range of $0.5 < z < 1.5$ and stellar masses $10^{10} < M/\msun < 10^{11}$ in GOODS-N field. The blue stars are GALEX (LBGs) from GOODS-N and the brown circles are from CDFS. The gray symbols are all galaxies from CDFS and the black triangles are galaxies from GOODS-N. The plots indicates that blue galaxies having lower surface density. \textit{Right panel}: The color ($z_{850} -  K$) versus surface density for galaxies at  $1.5 < z < 2.5$, as the green squares are the BM/BX galaxies from GOODS-N and the brown circles are pseudo BM/BX from CDF-S.} 
\label{fig8}
\end{figure*}

\begin{figure}
\includegraphics[width=3.4 in]{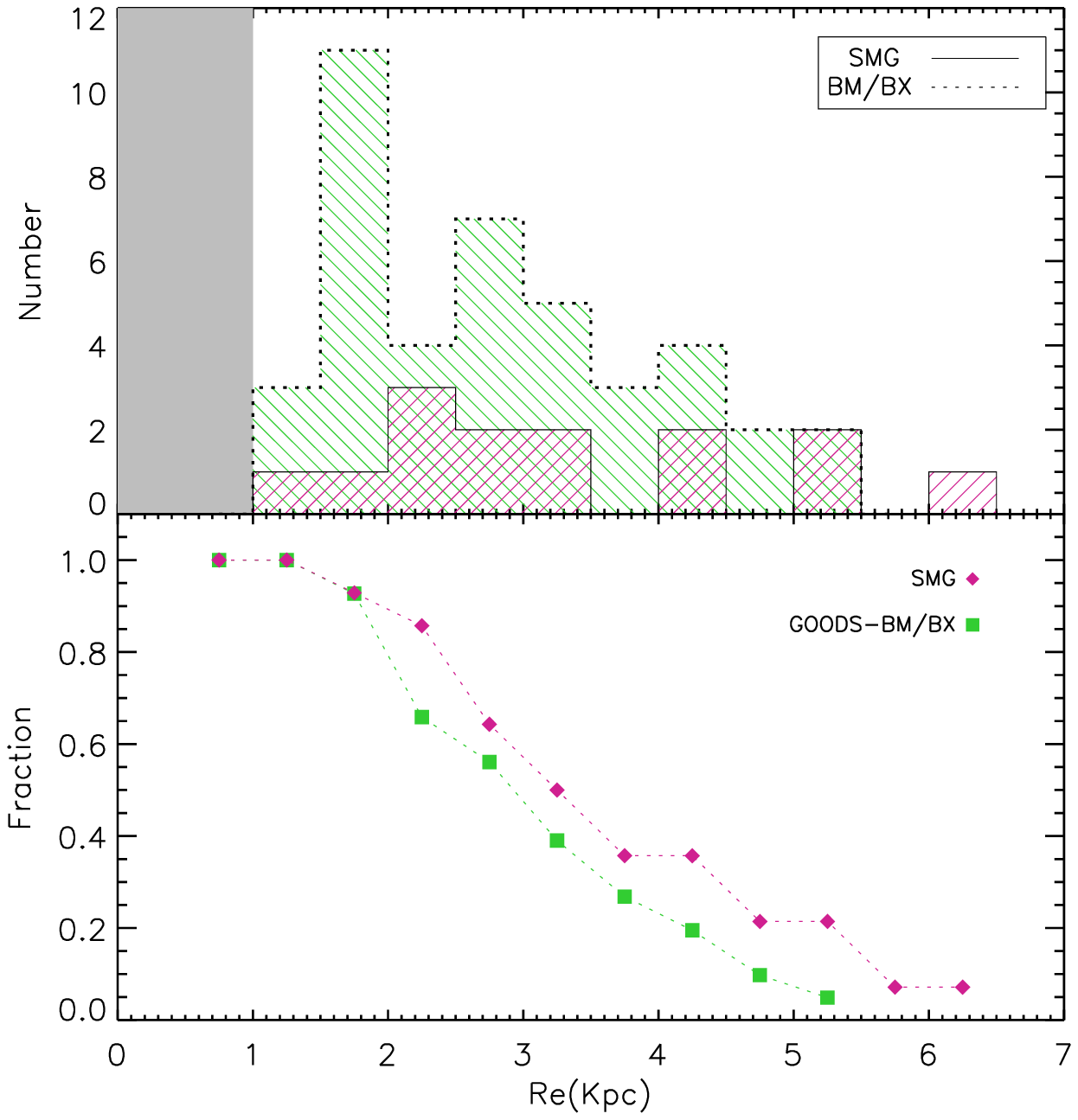}
\caption{ \textit{Top panel:} Histograms show the distribution of sizes of Submm galaxies (\textit{solid line}) and BM/BX galaxies (\textit{dotted lines}). The gray region indicates apparent effective radii below 0.1''($\sim$ 1 kpc at $z\sim2$) which size measurements have large uncertainties. \textit{Lower panel:} Cumulative distribution function for SMGs (pink dimonds) and BM/BX galaxies (green squares). As the plot shows, the distributions of half-light radii of these two populations are comparable with SMGs being slightly larger.}
\label{fig9}
\end{figure}

In Figure \ref{fig3} we show the size distributions for UV-bright galaxies between $10 < log(\mstar/\msun) < 11$. The blue histogram shows the measured half-light radii of GALEX/LBGs at $z \sim 1$  while the distribution of sizes measured for BM/BX galaxies ($z \sim 2$) are shown as a green histogram. It can be seen from this plot that UV-bright galaxies at $z \sim 2$ are smaller compared to similar galaxies at $z \sim 1$.  Specifically, the median half light radii of BM/BX galaxies in this mass range is $2.68 \pm 0.19$ kpc significantly smaller than the median effective radii of GALEX/LBGs ($4.42 \pm 0.52$ kpc). This means that sizes of UV-bright galaxies evolve by a median factor of $0.60\pm 0.08$ between $z \sim 2$ to $z \sim 1$. It is harder to measure the evolution for $z \sim 3$ to $z \sim 2$, as we have 4 LBGs within mass range of $10 < log(\mstar/\msun) < 11$  at $z \sim 3$ in our sample. The median half-light radius of these 4 massive galaxies is $2.22 \pm 0.61$ kpc, $0.82\pm 0.23$ smaller than $z\sim 2$ BM/BX galaxies. We perform a power-law fit on the size evolution and find $r_{e} \propto (1+z)^{-1.11\pm0.13}$ over the range $ 0.6 \lesssim z \lesssim 3.5$.
 
The size-redshift relation for star-forming galaxies with spectroscopic redshifts in the GOODS North field is shown in Figure \ref{fig4}. Galaxies are split into the four different stellar mass bins shown in the figure. In each panel, the gray small symbols are normal star forming galaxies with spectroscopic redshifts selected to have specific star-formation rates $log\ sSFR > -10$. Color symbols represent our samples of UV-bright galaxies and sub-millimetre galaxies (SMGs). LBGs ($2.7 < z < 3.5$) are shown as dark-green triangles and BM/BX galaxies ($1.4 < z < 2.7$) are plotted as green squares. GALEX/LBGs at relatively lower redshifts ($0.6 < z < 1.4$) are plotted in different panels as blue stars. The black symbols at $z \sim 0$ are star forming galaxies from \cite{guo2009} from SDSS.  

As Figure \ref{fig4} shows, sizes of UV-bright galaxies at a fixed stellar mass increase towards lower redshifts. There is also a trend that the size evolution maybe stronger for the most massive galaxies. For example, UV-bright galaxies with stellar masses $9 < log(\mstar/\msun) < 10$ evolve by a median factor of approximately $1.15 \pm 0.25$ from $z \sim 2$ to $z \sim 0.8$. However, galaxies with masses $10< log(\mstar/\msun) < 10.5$ grow by a median factor of approximately $1.82 \pm 0.25$. At higher stellar mass bin ($10.5 < log(\mstar/\msun) < 11$) this evolution is even stronger: a factor of $2.05 \pm 1.26$ over the same redshift range. However, selection effects may influence this trend (especially in the low-mass bin) and more complete samples are needed.

SMGs are also shown as pink diamonds in Figure \ref{fig4}; they are discussed further in section 6.4. Galaxies which have X-ray luminosities $ > 10^{42}$ $ergs $ $ s^{-1}$ (i.e. potential AGN hosts) are marked with open circles. These galaxies have sizes similar to the others. Since AGN is point like, one would expect galaxies with AGN emission to have smaller half-light radii. Their ``normal'' rest frame optical sizes are thus perhaps not strongly influenced. Nonetheless, due to possible effects on the size and mass estimates, X-ray detected galaxies should be considered low-confidence points.  

\subsection{Stellar Mass-Size Relation}

We now investigate how mass-size distribution of the UV-bright galaxies compares to other galaxy populations. We further check if the relation between stellar mass and size of these galaxies exists to high redshift. 

The stellar mass-size distributions of our samples are shown in Figure \ref{fig5}. We have divided our sample into three redshift bins, $0.5< z <1.5$, $1.5 < z < 2.5$, $2.5 < z < 3.5 $. UV-bright galaxies are color coded the same as Figure \ref{fig4}. Solid and dotted lines in each panel show the mass-size relation from \cite{shen2003} for star forming galaxies and early-type galaxies at $z \sim 0$, respectively. The gray symbols are the \textit{K}-selected sample of galaxies in CDF-South with stellar masses $ > 10^{9.8} \msun$ \citep{franx2008}. Comparing our results with the galaxies from CDF-S allows us to see where our galaxies lie relative to a purely mass-selected sample. Brown solid lines in the middle of each panel show the median effective radii in narrow mass bins for all galaxies in CDF-S. As can be seen, at $z\sim 1$ and $z\sim2$, UV-bright galaxies have larger effective radii compared to all galaxies at a given stellar mass. For galaxies between $10^{10}$-$10^{11} \msun$, the median sizes differ by a factor of $1.89\pm0.23$ and  $1.12\pm0.09$ at $z \sim 1$ and 2 respectively. As the spectroscopic sample contains only 4 UV-bright galaxies in this mass range at
$z\sim3$, the difference with CDF-S is not well constrained at this redshift ($1.27\pm0.36$).  

We further investigate how the size mass distribution of GOODS-N UV-bright galaxies compares to that of UV-bright galaxies in CDF-S. This would help to verify if the previously reported size-mass relation for a large photometric sample \citep{franx2008} is consistent with our spectroscopic sample. We define analogues of the $z \sim 1, 2$ and 3 LBGs in CDF-S by their optical colors and magnitudes. Specifically, we selected the analogues of the $z \sim 1$ GALEX/LBGs by requiring $(B - V) < 0.6$ and $R \leq 23.8$. This selection is based on the locus of the GOODS-N GALEX/LBGs in this color-magnitude plane. The analogues of BM/BX galaxies in CDF-S are similarly selected with $R < 25.1$ and $(B - V) < 0.8$ and $z_{phot} = 1.5-2.5$. We note that the $R$ band magnitude in both surveys is estimated by the average values of $V$ and $I$ band magnitudes. We verified that the spectroscopic samples are effectively selected by these criteria. In order to identify analogues UV-bright objects in CDF-S at $z \sim 2.5-3.5$, we find the location of LBGs in GOODS-N on the color-color $(B - V)$ versus $(V - I)$ diagram and apply the following criteria to the CDF-S galaxies; (1) $ 0.5 < (B - V) < 1.2 $ for $(V - I) < 0.35$ ;(2) $ (B - V) > 3 \times (V- I) + 0.5 $ and $(B - V) < 1.2 $ for $ 0.35 < (V - I) < 0.5$. We note that these selection criteria do not impose a ``U-drop out'' criteria, but do select galaxies with similar brightness and UV slope as the usual LBG methods.  

The comparison of size-mass distributions for UV-bright galaxies from both fields is shown in Figure \ref{fig6}. In each panel, the color symbols are the median log effective radii of UV-bright galaxies in narrow mass bins, with the blue and green symbols representing UV-bright galaxies in GOODS-N and the brown symbols are the ones in CDF-S. Error bars show the one $\sigma$ dispersion. UV-bright samples from both fields follow consistent size-mass relation, further confirming that UV-bright galaxies are on average larger than overall galaxies at a fixed stellar mass. We note that at high redshift ($z \sim 2.5-3.5$) the comparison is weak because of the limited number LBGs in GOODS-N.

The UV-bright galaxies show a weak stellar mass size relation at $z = 1$ and $z = 2$ in Figure \ref{fig5}. To quantify this relation, we fitted a power law function of the form $r_{e} \propto M^{\alpha}$ to the individual UV-bright galaxies in each redshift bin. The fitting results are plotted as color dashed-dotted lines in Figure \ref{fig6}. At $z \sim 1$, the size of GALEX/LBGs in GOODS-N scales with stellar mass as $r_{e} \propto M^{0.19\pm0.05}$ and the BM/BX galaxies at $z \sim 2$ have $r_{e} \propto M^{0.30\pm0.06}$. The uncertainties were estimated using bootstrap resampling. The results are comparable to the relation for late-type galaxies at $z \sim 0$ from \cite{shen2003}. They found $r_{e} \propto M^{0.15}$ for low-mass ($log(M) < 10.6$) late-type galaxies and steeper relation for high mass late type galaxies ($r_{e} \propto M^{0.4}$). In the highest redshift bin ($z \sim 3$), we use UV-bright galaxies from CDF-S to find the best fit ($\alpha = 0.32\pm0.06$). Table \ref{tb2} lists the best-fit power law parameter to the mass size relation of UV-bright galaxies. The slopes of the mass-size relation at different redshift bins are consistent and there is no significant evolution. Our results confirm the persistence of the size-mass relation for star forming galaxies up to high redshift. 

\begin{table}\small 
\caption{Best fits power law parameter for the stellar mass - size relation}
\centering
\label{tb2}
\begin{tabular*}{3.4 in}{c c c}
  \hline
 \hline
 Sample          & Redshift          &  $\alpha$       \\  \hline
 GALEX/LBG       & $0.6 < z < 1.4$   &  $0.19\pm0.05$  \\
 BM/BX           & $1.4 < z < 2.7$   &  $0.30\pm0.06$  \\
 CDFs-UV Bright  & $2.5 < z < 3.5$   &  $0.32\pm0.06$  \\
 \textit{sBzK}   & $1.4 < z < 3.2$   &  $0.31\pm0.09$  \\
 \hline 
\end{tabular*}
\footnotetext{\textbf{Note}. Power law parameter $\alpha$ is defined as $r_{e} \propto M^{\alpha}$.}
\end{table}

It is worth checking whether or not other star forming populations at high redshift have the same size-mass relation. Therefore, we compare star forming \textit{BzK} galaxies (\textit{sBzK}) with BM/BX galaxies. The size distribution as a function of stellar mass is shown in Figure \ref{fig7} with the \textit{sBzK} galaxies plotted as gray filled circles. The median of log effective radii of \textit{sBzK} galaxies are overplotted as red squares. BM/BX galaxies and SMGs are marked with the green and pink circles respectively and the green dashed-dotted line shows the mass-size relation for BM/BX galaxies. As one can see, the mass-size relation for \textit{sBzK} galaxies is comparable to the one for BM/BX galaxies. The average offset of median half light radii of  \textit{sBzK} galaxies to the BM/BX mass-size relation is 0.037 dex. The effective radii of \textit{sBzK} galaxies scale with stellar mass as $r_{e} \propto M^{0.31\pm0.09}$ close to the size-mass relation for UV-bright galaxies. This similarity could be due to the significant overlap between the UV-selected galaxies and \textit{sBzK} galaxies \citep[e.g.][]{reddy2005}. We note that stellar masses can have significant uncertainties depending on the methods used to calculate them. We verify that by using the stellar mass estimates from \cite{reddy2006} for UV-bright galaxies, the size-mass distribution of these galaxies covers the same region in the size-mass plane (see e.g., middle panel of Figure \ref{fig5}). Hence, this shows that our results are robust against using different estimates of stellar masses. 

\subsection{Color - Surface Density}

One of the surprising results of \cite{franx2008} was the tight correlation between stellar mass surface density ($\mstar/r_{e}^{2}$) and color of galaxies to $z \sim 3$. They showed that bluer galaxies have lower surface densities than red quiescent ones. They indicated that the color of galaxies correlates more fundamentally with the stellar mass surface density than the mass. 

We show the tight relation between color and stellar mass surface density for both star-forming and quiescent galaxies with $\msun \sim 10^{10}-10^{11}$ in Figure \ref{fig8}. In the left panel the observed color ($V_{606} - z_{850}$) is plotted versus the stellar mass surface density for galaxies at redshift $\sim 1$. The black and gray symbols are galaxies from GOODS-N and CDF-S fields, respectively. The blue stars and brown circles indicate the GALEX (LBGs) and their analogues from CDFS respectively. In the right panel, the relation between observed ($z_{850} - K$) and surface density for galaxies at higher redshift $\sim 2$ is illustrated , with BM/BX galaxies in GOODS-N shown by green squares and their analogues in CDF-S as brown circles. As can be seen in this Figure, the blue star forming galaxies have lower surface density than red galaxies at both $z \sim 1$ and $z \sim 2$, though the correlation between color and surface density is more tight at $z \sim 1$ than $z \sim 2$. In general, using sample of galaxies with spectroscopic redshifts confirms that the color of galaxies are tightly correlates with surface density and this relation holds out to high redshifts. 
 
\subsection{Sizes of Submm Galaxies}
Since GOODS-N includes 14 SMGs with redshifts, we compare their mass and sizes to the UV-selected galaxies at $z \sim 2$. As Figure \ref{fig4} illustrates, the optical rest-frame sizes of the SMGs (pink diamonds) and UV-selected galaxies of similar mass are comparable. The median half light radii of all SMGs is $2.90 \pm 0.45$ kpc which is in good agreement with the median optical rest-frame sizes of the BM/BX galaxies over the whole mass range ($2.68 \pm 0.25$ kpc). The median effective radii of SMGs with stellar masses of $10^{10}-10^{11}$ is $2.65 \pm 0.56$ which is similar to the BM/BX of at the same mass range ($2.68 \pm 0.19$). The errors are computed using bootstrap resampling. 

We further compare the size distributions of sub-mm galaxies and BM/BX galaxies in Figure \ref{fig9}. In the top panel, the histograms show the size distribution of BM/BX galaxies and SMGs. In the bottom panel, the normalized cumulative distribution functions are shown for both samples. As this plot shows, the size distributions of SMGs are comparable with $z \sim 2$ UV-bright galaxies and the SMGs follow the same size distribution as BM/BX galaxies in the optical rest-frame with SMGs being slightly larger. According to a KS test, the probability that two distributions being the same is 72 percent, suggesting no significant difference between the rest-frame optical sizes of the SMGs and UV-bright galaxies.

\section{Discussion}

\subsection{The growth of star-forming galaxies}

\begin{figure}
\includegraphics[width=3.4 in]{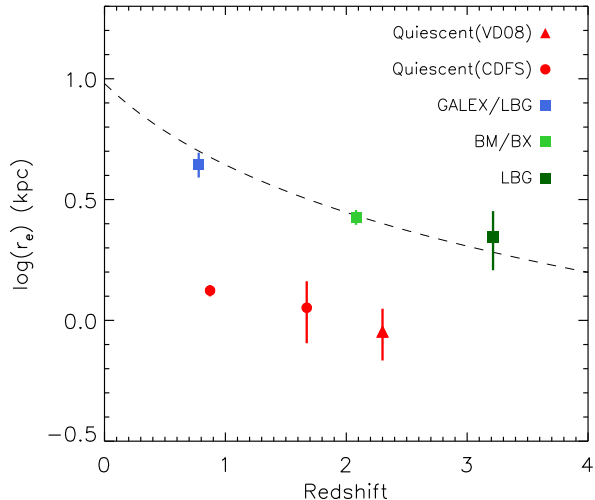}
\caption{The median sizes of UV-bright galaxies (squares) as a function of redshifts for galaxies with stellar masses $10^{10} < M/\msun < 10^{11}$ in GOODS-N field. The red filled circles are quiescent galaxies from CDF-S study with similar mass range and red triangle is quiescent sample from  \cite{vandokkum2008} with median stellar masses of $1.7 \times 10^{11} \msun$. The dashed line shows the best fitting size evolution to the UV-bright galaxies ($r_{e} \propto (1+z)^{-1.11\pm0.13}$). The plot indicates that the UV-selected galaxies are larger than quiescent galaxies and their sizes evolve with redshift.}
\label{fig10}
\end{figure}

We study the size evolution of galaxies with spectroscopic redshifts between $z \sim 0.5-3.5$. Our results are summarized in Figure \ref{fig10}. As this plot illustrates, at each epoch, the median sizes of UV-bright galaxies (color squares) with stellar masses $10 < log(\mstar/\msun) < 11$  are larger (by a median factor of $0.45\pm0.09$) than quiescent galaxies in a similar mass range selected from CDF-S (red circles). The red triangle is the median sizes of quiescent galaxies studied in \cite{vandokkum2008} (VD08 hearafter) with a median stellar mass of $1.7 \times 10^{11} \msun$ and median redshift of 2.3. Table \ref{tb3} lists the median half light radii for UV-bright and quiescent galaxies seen in Figure \ref{fig10}. The growth of UV-bright galaxies with time is also illustrated in this figure. The dashed line shows that UV-bright galaxies scaled up their sizes towards lower redshifts as $(1+z)^{-1.11\pm0.13}$. 

Previous studies \citep{franx2008, Kauffmann2003} reported that there is a correlation between color and stellar mass surface density of galaxies at both low and high redshifts. However, the results based on photometric redshifts can be uncertain especially for star forming galaxies. By means of our large sample of galaxies with secure redshifts, we have confirmed that the tight correlation between color and stellar mass surface density of galaxies that holds out to high redshifts; as the stellar mass densities of blue star forming galaxies are smaller than those of red quiescent ones. This verifies that the galaxies with higher specific star formation rate are larger than the ones with lower specific star formation rates.

\begin{table}\footnotesize
\caption{Median sizes of different samples}
\centering
\label{tb3}
\begin{tabular}{c c c c}
  \hline
 \hline
   Sample          & Redshift         &   Size           & $ median(log(\mstar/\msun)$  \\  \hline
   GALEX/LBG       & $0.6 < z < 1.4$  & $4.42 \pm 0.52$  & $10.2\pm0.3$ \\
   BM/BX           & $1.4 < z < 2.7$  & $2.68 \pm 0.19$  & $10.4\pm0.2$ \\
   LBG             & $2.7 < z < 3.5$  & $2.22 \pm 0.61$  & $10.6\pm0.2$ \\
   CDFS            & $0.5 < z < 1.5$  & $ 2.33 \pm 0.1 $ & $10.4\pm0.3$ \\
   CDFS            & $1.5 < z < 2.5$  & $ 2.40 \pm 0.13$ & $10.4\pm0.3$ \\
   CDFS            & $2.5 < z < 3.5$  & $ 1.73 \pm 0.18$ & $10.4\pm0.3$ \\
   Quiescent(CDFS) & $0.5 < z < 1.5$  & $1.32 \pm 0.07$  & $10.5\pm0.3$\\
   Quiescent(CDFS) & $1.5 < z < 2 $   & $ 1.12 \pm 0.32$ & $10.7\pm0.2$\\
   Quiescent(VD08) & $2 < z < 2.5 $   & $ 0.9 \pm 0.21 $ & $11.23$ \\
   SMG             & $0.5 < z < 3.$   & $ 2.90 \pm 0.45$ & $10.8\pm0.5$ \\

 \hline 
 
 \end{tabular}
\footnotetext{\textbf{Notes.} The median sizes of samples are for galaxies with stellar masses between $10^{10}-10^{11}\msun$ except for the quiescent(VD08) \citep{vandokkum2008} and SMG galaxies. The CDFS sample is all galaxies at the same stellar mass range in CDF-S field.}
\end{table}

We have also explored the stellar mass-size distribution for galaxies in our sample. We have confirmed that there is a relation between stellar mass and size for UV-bright galaxies at $0.5 < z <3.5$. The relations are consistent for both spectroscopic and photometric redshift samples. Our results verify that there is no significant evolution of the slope of the size-mass relation to redshift $\sim 3.5$. The evolution of the relation from $z=1$ to $z=0$ is not well established, as it is not straight forward to select Lyman Break galaxies at $z\sim0$. Some authors find evolution \citep[e.g.][]{franx2008, williams2009} in the stellar mass-size relation, however, \cite{barden2005} have reported that the stellar mass-size relation for disk galaxies remains constant from $z\sim 1$ to the present. Selection effects likely play a role. Comparing to the theoretical work, the size evolution predicted by \cite{somerville2008} is somewhat stronger than the evolution we measured here.  

Star forming galaxies at high redshifts can be identified by means of different methods of sample selection and can be assigned as different populations of galaxies (e.g, sub-mm or \textit{sBzK}) and therefore studied separately. However, many of these galaxy populations may overlap, hence it worth checking out if they have comparable properties. We therefore, compare sizes of UV-bright galaxies to other types of star-forming galaxy populations at high redshift, i.e. \textit{sBzK} galaxies and SMGs. We show that these two populations have comparable sizes to the UV-bright galaxies. This suggests that in general, star forming galaxies have larger sizes than quiescent ones without regard to their type. However, sample selection might alter the results. We return to this possibility later, in the section 7.3.

\subsection{Comparison to previous work}
The strong growth of both star forming and quiescent galaxies has been shown by several authors  \citep{trujillo2006a, franx2008, williams2009, toft2007}. The size growth of star forming galaxies studied here going as $(1+z)^{\alpha}$ where $\alpha =-1.11\pm0.13$, agrees well compared to previous studies where the size evolves with $\alpha \sim -1$. For example, \cite{bouwens2006} found $\alpha = -1.1 \pm 0.3$ for a sample of luminosity-selected dropout galaxies between $z \sim 6$ and $z \sim 2.5$ while \cite{Dahlen2007} find  $\alpha = -1.10 \pm 0.07$ for luminosity-selected disk galaxies over the range $z \sim 1.1-2.2$. \cite{williams2009} also find a similar slope for star forming galaxies in UDS at $z \sim 0.5-2$.

The similarity between sizes of UV-bright galaxies and star forming \textit{BzK} galaxies is also shown in our study. The two populations are also compared by \cite{overzier2010}. They showed that rest-frame optical sizes of \textit{sBzK} galaxies are somewhat larger than UV-selected galaxies. Since \textit{BzK} galaxies are K-selected and thus tend to represent massive galaxies, it is not surprising that they appear larger. Hence, the sizes must be corrected for differing stellar masses for an accurate comparison to be made. 

Our results also show that the SMGs have a median half light radii of $2.90 \pm 0.45$ kpc comparable to the rest-frame optical sizes of the BM/BX galaxies. The result is in agreement with a recent study by \cite{swinbank2010} where they use deep \emph{HST} \emph{I} and \emph{H}-band imaging and show that the rest-frame optical sizes of the SMGs and UV-bright galaxies in GOODS-N are comparable. The typical half-light radii of the SMGs in the \emph{H}-band (rest-frame optical) measured by \cite{swinbank2010} is $2.8 \pm 0.4$ which is consistent with our measurements. \cite{almaini2005} also compared sizes of submm galaxies with Lyman-break galaxies at rest-frame optical and found no clear difference between sizes of the SMGs and LBGs. It is worth noting that SMGs have larger median stellar masses compared to BM/BX galaxies (also mentioned in \cite{swinbank2010}). This suggests higher stellar mass densities for SMGs compared to UV bright galaxies. However, the similarity in sizes between these two galaxy populations at high $z$ does not by itself allow conclusive connections to be drawn between the two populations. The SMGs that are faint or lacking emission lines could be missing from our spectroscopic sample and they may have dramatically different mass or size properties. Therefore, selection effect could bias these results. Spectroscopic observation in near-IR might help to solve this problem; e.g., \cite{kriek2009b} find a median size of 2.8 kpc for a small sample of near-IR spectroscopically confirmed star forming galaxies at $z\sim2.3$. Indeed, our measured size evolution is consistent with their spectroscopic sample; however, near-IR spectroscopy over a large redshift range is needed to definitively quantify the size evolution.

\subsection{caveats}

With this large spectroscopic sample of high redshift galaxies, we have removed an important source of uncertainty in the stellar mass size relation and verified the results from larger samples based on photometric redshifts. However, the spectroscopic catalog we used in this study consists of many different spectroscopic surveys which made it inhomogeneous. Galaxies in this catalog are mostly relatively unobscured star forming galaxies with emission lines, comparing these to quiescent galaxies can still be a problem because spectroscopy of high redshift red galaxies is difficult. Therefore, using a larger and homogeneous sample of high redshift galaxies with secure redshift will allow greatly improved constraints on their evolution. New ground and space-based IR multi-object spectrographs (WFC3 grism, MMIRS, MOSFIRE) will allow investigations of more complete samples of galaxies and studies the correlation between galaxy properties at higher redshifts. Dust-obscured starbursts are also likely to be missed in our UV-bright sample, hence NIR spectroscopy will also permit studies of this important population. \\

\section{Summary}
We have performed the first size evolution study of UV-bright galaxies with secure redshifts in the GOODS-North field. We derive half light radii of galaxies at a wide range of redshifts (up to $z \sim 3.5$). Using sample of galaxies with secure redshifts, we have quantified the size evolution of galaxies without the potential uncertainty of photometric redshifts and confirm the previous studies  that galaxies of similar mass were generally smaller at past and their sizes evolve with redshift. Specifically, we find that:

\begin{itemize}
\item UV-bright galaxies evolve strongly as $(1+z)^{-1.11\pm0.13}$ confirming previous studies based on photometric redshifts \citep[e.g.][]{trujillo2006a, trujillo2007, franx2008, williams2009}.
\item UV-bright galaxies are significantly larger than quiescent galaxies. At the same mass, the median difference is $0.45\pm0.09$ dex.
\item The LBG, BM/BX and GALEX/LBG samples show smooth evolution with redshift, indicating that these techniques indeed select similar UV-bright galaxies at different redshifts.
\item The SMGs have half-light radii similar to UV-bright galaxies of the same mass.

\end{itemize}

\begin{acknowledgments}
We thank anonymous referee for helpful comments and suggestions. We thank Rychard Bouwens for providing us $K$-band images. MM is supported by the Marie Curie Initial Training Network ELIXIR (Early unIverse eXploration with nIRspec), grant agreemnet PITN-GA-2008-214227 (form the European Commission). RJW acknowledges support from the Nederlandse Onderzoekschool voor Astronomie (NOVA) and NSF grant AST-0707417.
\end{acknowledgments}


\end{document}